\documentclass{article}

 \usepackage[dblblindworkshop,final]{neurips_2025}
 \usepackage{graphicx}
 \usepackage{amsmath}
 \usepackage{amssymb}

\usepackage[utf8]{inputenc} % allow utf-8 input
\usepackage[T1]{fontenc}    % use 8-bit T1 fonts
\usepackage{hyperref}       % hyperlinks
\usepackage{url}            % simple URL typesetting
\usepackage{booktabs}       % professional-quality tables
\usepackage{amsfonts}       % blackboard math symbols
\usepackage{nicefrac}       % compact symbols for 1/2, etc.
\usepackage{microtype}      % microtypography
\usepackage{xcolor}         % colors

\title{Toward Complete Merger Identification at Cosmic Noon with Deep Learning}
\workshoptitle{Machine Learning and the Physical Sciences (ML4PS)}

\author{%
  Aimee Schechter \\
  Department of Astrophysical and Planetary Sciences\\
  University of Colorado Boulder\\
  Boulder, CO 80309 \\
  \texttt{aimee.schechter@colorado.edu} \\
  \AND
  Aleksandra \'Ciprijanovi\'c \\
  Computational Science and AI Directorate\\ 
  Fermi National Accelerator Laboratory\\ 
  Batavia, IL 60510 \\
  Department of Astronomy and Astrophysics\\
  University of Chicago\\ 
  Chicago, IL 60637\\
  \texttt{aleksand@fnal.gov}
  \And
   Rebecca Nevin\\
   Computational Science and AI Directorate\\ 
  Fermi National Accelerator Laboratory\\ 
  Batavia, IL 60510 \\
  \texttt{rnevin@fnal.gov}
  \And
   Julie Comerford \\
  Department of Astrophysical \\and Planetary Sciences\\
  University of Colorado Boulder\\
  Boulder, CO 80309 \\
  \texttt{julie.comerford@colorado.edu}
  \And
  Xuejian Shen\\
TAPIR, California Institute of Technology \\
Pasadena, CA 91106\\
Kavli Institute for Astrophysics and Space Research\\
Massachusetts Institute of Technology\\
Cambridge, MA 02139\\
  \texttt{xuejian@mit.edu}
  \And  Aaron Stemo\\
  Department of Physics and Astronomy\\
  Vanderbilt University\\ 
  Nashville, TN 37235 \\
  \texttt{aaron.m.stemo@vanderbilt.edu}
    \And Laura Blecha\\
Department of Physics\\
University of Florida\\ 
Gainesville, FL 32611 \\
  \texttt{lblecha@ufl.edu}
}

\begin{document}

\maketitle

\begin{abstract}
As we enter the era of large imaging surveys such as \textit{Roman}, \textit{Rubin}, and \textit{Euclid}, a deeper understanding of potential biases and selection effects in optical astronomical catalogs created with the use of ML-based methods is paramount. 
This work focuses on a deeper understanding of the performance and limitations of deep learning-based classifiers as tools for galaxy merger identification.
We train a ResNet18 model on mock Hubble Space Telescope CANDELS images from the IllustrisTNG50 simulation. 
Our focus is on a more challenging classification of galaxy mergers and nonmergers at higher redshifts $1<z<1.5$, including minor mergers and lower mass galaxies down to the stellar mass of $10^8 M_\odot$. 
We demonstrate, for the first time, that a deep learning model, such as the one developed in this work, can successfully identify even minor and low mass mergers even at these redshifts.
Our model achieves overall accuracy, purity, and completeness of 73\%. 
We show that some galaxy mergers can only be identified from certain observation angles, leading to a potential upper limit in overall accuracy.
Using Grad-CAMs and UMAPs, we more deeply examine the performance and observe a visible gradient in the latent space with stellar mass and specific star formation rate, but no visible gradient with merger mass ratio or merger stage. 
\end{abstract}

\section{Introduction}
\label{sec:introduction}

Galaxies grow through cosmic time hierarchically.
%with smaller structures merging into larger ones.
In addition to growing in total mass, this process also forms new stars, changes galaxy morphologies, and can trigger active galactic nucleus (AGN) activity (e.g., \cite{barnes_fueling_1991, mihos_gasdynamics_1996, di_matteo_energy_2005, springel_black_2005, springel_formation_2005, springel_modelling_2005, ellison_galaxy_2008, patton_galaxy_2011, barrows_census_2023}).
To fully understand the role of galaxy mergers in star formation and AGN activity, we need large catalogs of merging galaxies at different merger stages and across a range of stellar masses and merger mass ratios. 
However, this can be challenging as many non-parametric methods are calibrated at low-$z$ and are targeted at identifying major mergers (merger mass ratio $\mu \geq$ 1/4) among high mass galaxies ($M_\star \gtrsim 10^{10}M_\odot$). 
These methods quantify the distributions of light in an image to measure how concentrated, clumpy, or asymmetric the distribution is \cite{conselice_relationship_2003, lotz_new_2004}. 
The merger stage adds another complication. 
Finding early-stage mergers with two identifiable nuclei and faint features like tidal tails is easier for non-parametric methods (e.g., tidal tails causing asymmetry) than finding late-stage mergers near coalescence. 
Close pair analyses, which find galaxies that are within a given separation in physical and velocity space, can identify early-stage mergers, but only looking at early stages leaves out half of the merging population.
Combining non-parametric methods through Linear Discriminant Analysis (LDA) or a Random Forest (RF) (e.g., \cite{nevin_accurate_2019, snyder_automated_2019, rose_identifying_2023, wilkinson_limitations_2024}) is one way to get a more accurate and diverse catalog of mergers.

Convolutional Neural Networks (CNNs) offer an even more flexible method for finding mergers at different merger stages and redshifts since they have the ability to utilize all features present in galaxy images. 
They have already been applied to multiple mock and real imaging survey datasets (e.g., \cite{bottrell_deep_2019,ciprijanovic_deepmerge_2020, bickley_convolutional_2021, margalef-bentabol_galaxy_2024, rose_ceers_2024}). 
Still, the majority of these studies are focused on lower redshifts (all are at $z < 1$ except \cite{ciprijanovic_deepastrouda_2023} at $z = 2$ and \cite{rose_ceers_2024} at $3 < z < 5$), and higher mass galaxies (all above $M_\star = 10^9 M_\odot$ except \cite{rose_ceers_2024}). 
We apply our CNN to a sample that includes both higher-$z$ galaxies ($ 1 < z < 1.5$) and lower mass galaxies ($M_\star > 10^8 M_\odot$).
Our aim is that by using machine learning (ML) rather than visual identification, we can avoid biases of identifying only more obvious mergers.
%, such as a merger between two large spiral galaxies.
Additionally, CNNs are not restricted to any redshift or mass range, and could potentially be able to identify even less visible merger features (such as those in minor mergers or high-$z$ galaxies).
We aim to use interpretive tools, such as examining important regions of the image with Gradient-weighted Class Activation Mapping \cite[Grad-CAM;][]{selvaraju_grad-cam_2020} and the latent space with UMAPs~\cite{mcinnes_umap_2020}, to better understand how to identify high-$z$ mergers and why our network made its decisions.

\section{Data}
\label{sec:data}
Cosmological simulations have proven to be useful tools in training ML algorithms for merger identification since there is a ground truth that is separate from any other merger identification tools, including by-eye classification. 
This work uses the IllustrisTNG cosmological magnetohydrodynamical simulation suite \cite{pillepich_first_2019, nelson_first_2019}.
We use the smallest TNG50 box with 50 comoving Mpc per side.
Its $\sim0.1\,\mathrm{kpc}$ spatial resolution and $\sim8\times10^4M_\odot$ baryonic mass resolution provide a diverse set of galaxy morphologies, stellar masses, and details that may be important for distinguishing mergers from nonmergers such as star-forming clumps. 
We utilize the definition of a merger from \cite{rodriguez-gomez_merger_2015} and apply a minimum mass cutoff: any subhalo (galaxy) at least 1000 times the baryonic mass resolution with two direct progenitors in the previous time snapshot is classified as a merger.
We use two full snapshots (which include full physics outputs necessary to run radiative transfer): $z = 1$ and $z = 1.5$.
We apply a 500Myr time window centered around those snapshots, and anything that merges at any point in that window is considered a merger. All images are taken at the two central snapshots, which means that our sample includes early-stage mergers that merge later in the window but have not yet merged at the central snapshot, and late-stage mergers that merge early in the window and are near coalescence at the central snapshot.
For each merging galaxy found, we find a corresponding, mass-matched non-merging galaxy in the same snapshot.

To create our mock images, we first use the radiative transfer code SKIRT that includes dust and AGN~\cite[version 9;][]{baes_efficient_2011,baes_skirt_2015,camps_skirt_2015,camps_skirt_2020} (for details see \cite{vogelsberger_high-redshift_2020} and \cite{shen_high-redshift_2020, shen_high-redshift_2022}).
Each extracted galaxy is observed from six viewpoints.
We then filter the wavelengths down to those of the \textit{HST} CANDELS F814W, F160W, and F606W filters~\cite{koekemoer_candels_2011} for our three-channel input.
Additionally, we rebin to the camera's pixel scale and convolve with the PSF from the Tiny Tim software \cite{krist_20_2011} in that filter (following\cite{nevin_accurate_2019}).
The CNN must be able to distinguish between merging galaxies and background sources, so we place our mock galaxies in realistic environments by creating cutouts from real CANDELS mosaics that are not centered on any sources.
These cutouts also introduce background noise that gives a $5\sigma$ limiting magnitude of 26.5.
Any galaxy that had issues producing a reliable radiatively transferred image was thrown out, but we kept the matched counterpart because it did not drastically change the balance of the dataset.

Our dataset is split into training (70\%), validation (15\%), and test (15\%) sets.
All viewpoints of any given galaxy are in the same set.
The images are normalized between 0 and 1 with a log stretch.
The training set includes data augmentation through rotation up to $30^{\circ}$ and a vertical or horizontal flip, which makes the total number of images in the training set 5940 mergers and 5916 nonmergers, the validation set 630 mergers and 624 nonmergers, and the test set 630 mergers and 630 nonmergers. 
Example images are shown in the bottom right of  Figure~\ref{fig:figure}.

\section{Methods}
\label{sec:methods}

We use the ResNet18 architecture \cite{he_deep_2015}, with pre-trained weights from Zoobot2.0.2 \cite{walmsley_zoobot_2023} in PyTorch. 
We set the initial learning rate $10^{-5}$, and employ an exponential learning rate decay of 0.5 and cross-entropy loss with the Adam optimizer.
This is a binary classification (merger or nonmerger), so after convolutional layers, we change the head to have 2 output nodes. 
Early stopping is triggered at epoch 48 when the validation set loss does not improve by at least 0.0005 for 5 epochs.
The epoch with the lowest validation loss (epoch 43) is used to save the best model\footnote{Data and code are available at \url{https://github.com/alschechter/NeurIPSCosmicNoonMergerID}.}.

We examine the performance of our model using standard metrics such as accuracy, completeness, and purity. 
To further examine the behavior of our trained model, we utilize GradCAMs~\cite{selvaraju_grad-cam_2020}, which use the gradients heading into the final convolutional layer of a network to identify the key pixels for a given class.
UMAP~\cite[][]{mcinnes_umap_2020} is a dimensionality reduction technique, which we use to examine the high-dimensional latent space of our model (penultimate layer).
By seeing how close different galaxies in our test set appear on a UMAP, we can determine what physical processes the CNN may have recognized.

\section{Results}
\label{sec:results}

\begin{table*}[]
    \centering
    \begin{tabular}{c|c|c|c|c|c}
        Accuracy & Purity & Completeness & Brier Score & ECE & AUC\\
        \midrule
         $73.0\pm{0.4}\%$  &
         $74.0\pm{0.01}\%$ & $72.0\pm{0.01}\%$ & $0.19\pm{0.01}$ & $0.08\pm{0.03}$ & $0.8\pm{0.01}$\\
    \end{tabular}
    \caption{Mean and standard deviation of Accuracy, Purity, Completeness, Brier Score, ECE, and AUC for our model's three random seeds. }
    \label{tab:pcbe}
\end{table*}

\begin{table*}[]
    \centering
    \begin{tabular}{*{6}c}
    \multicolumn{6}{c}{Accuracy} \\
    \midrule
     All Mergers & Major & Minor  & Early Stage  & Late Stage  & Nonmergers\\
    \midrule
         $71.9\pm{1.0}\%$ & $75.8\pm{0.8}\%$ & $68.3\pm{0.7}\%$  & $79.6\pm{0.7}\%$ & $66.0\pm{0.01}\%$  & $74.0\pm{0.01}\%$  \\

    \end{tabular}
    \caption{Mean and standard deviation of our model's accuracy in three random seeds broken down by different subsets of galaxies.}

    \label{tab:acc}
\end{table*}

\begin{figure*}[!ht]
    \centering
    \includegraphics[width=\textwidth]{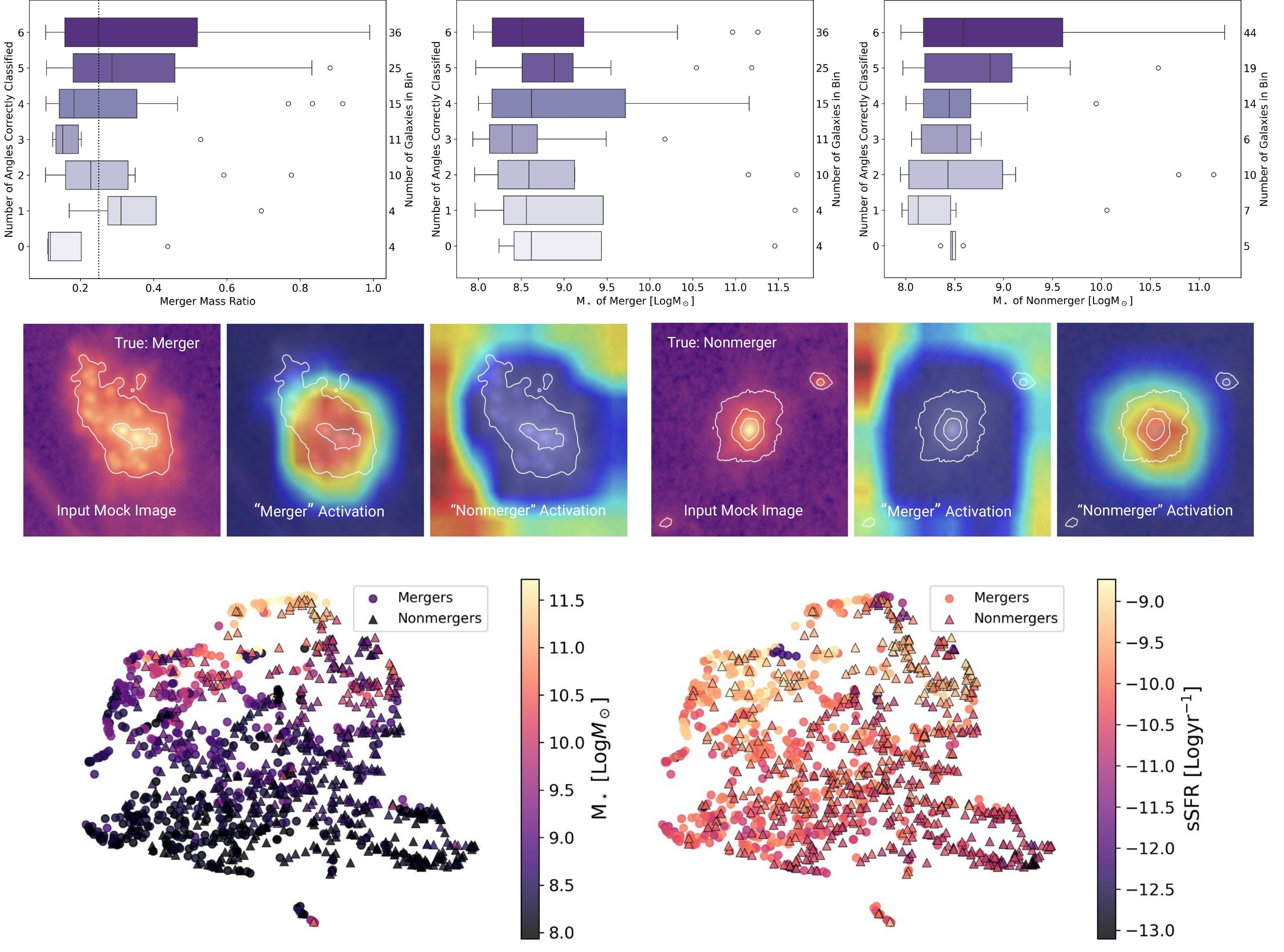}
    \caption{\emph{Top row:} Box plots showing the 25th-75th percentile (inter-quartile range, IQR) of the data inside the box, with the whiskers stretching to the furthest point within 1.5 times the IQR. The remaining points are outside of that range. These show the overall distributions of galaxies by merger mass ratio (\emph{left}; dotted line separates major and minor mergers) and stellar masses (\emph{center and right}), which are identified correctly from a given number of observing angles. \emph{Middle row}: Grad-CAM showing that the network focuses on the central galaxy when making its prediction. \emph{Bottom row}: UMAPs showing the true class by shape (nonmergers triangle, mergers circle) and colored by stellar mass (\emph{left}) and sSFR (\emph{right}), respectively, which show a clear gradient.}
    \label{fig:figure}
\end{figure*}

We train three models with different random seed initializations to ensure model stability. 
The mean accuracy, completeness, and purity of all three random seeds are shown in Table~\ref{tab:pcbe}.
All plots are from Seed 626, as it has both the highest accuracy and completeness.
All seeds in our model classifying galaxies with $M_\star > 10^8M_\odot $ at $1<z<1.5$ have accuracy of $\sim73\%$, similar to models reported for galaxies at $M_\star > 10^{9}$ at $0.1 < z < 1$ in\cite{margalef-bentabol_galaxy_2024}.
We examine Grad-CAMs, UMAPs, and the effect of observation angle to dig into what may have been causing difficulties in classifying galaxies.

The Grad-CAMs (Figure~\ref{fig:figure}, bottom left) show that the network focuses on the galaxy when it makes its decision. The central galaxy is highlighted when activating the predicted class, and the edges are highlighted for the non-predicted class. 
This is true for mergers and nonmergers.
The Grad-CAMs are not different enough between classes to draw conclusions about specific image features, but do offer confidence that the network has learned not to focus on background noise or sources, an important quality of successful merger identification using CNNs \cite{bottrell_deep_2019}.

We examine UMAPs of the test set images in terms of multiple physical quantities of the galaxies. 
The mergers (circles) primarily live on the bottom side of the UMAP, and nonmergers (triangles) on the top. 
However, there is a lot of overlap in the middle.
We see a clear gradient in the UMAP when colored by the stellar mass of each galaxy (Figure~\ref{fig:figure}, bottom center). 
The low-mass galaxies are on the bottom of the UMAP, with stellar mass increasing towards the top.
No stellar mass information was provided during training, but the network was able to recognize this physically meaningful quantity.
There is again a clear gradient in the UMAP when colored by specific star formation rate (sSFR; Figure~\ref{fig:figure}, bottom right).
The less star-forming galaxies are on the bottom right with sSFR increasing towards the upper left, even though no sSFR information was input to the model.
Because of this gradient, we speculate some of the nonmergers misclassified as mergers may be due to high, clumpy star formation, likely in the nonmergers seen in the top left.
There were no obvious trends for UMAPs relative to the merger stage or merger mass ratio.

There may be a ceiling of around 85\% accuracy for merger identification due to some morphological disturbances not being visible from every observation angle, and some clumpy nonmergers being indistinguishable from minor mergers \cite{bickley_effect_2024}.
Each galaxy in our test set is observed from six different angles. We examine the number of viewing angles from which a given galaxy is correctly identified, as a function of merger mass ratio and stellar mass, using box plots.
On the top left panel of Figure~\ref{fig:figure}, we can see that almost all mergers were identified correctly from at least one angle. 
Major mergers ($\mu \geq 1/4) $ can cause large morphological disruptions, and thus we expect them to be easier to identify than minor mergers $(1 / 10 < \mu < 1/4)$.
Overall, as the mass ratio increases, the merger is identified correctly more often.
However, we note that not all major mergers are correctly identified from more than three viewpoints.
The misclassification of major mergers could be due to a major merger between lower-mass galaxies, and thus it is harder to identify than a merger between high-mass galaxies.
Alternatively, if one of the galaxies is large and spheroidal, it could be blocking the companion galaxy, making it invisible from some angles.
Finally, a late-stage merger can be tricky to identify even among major mergers, but CNNs have been proven to be capable of it \cite{bickley_convolutional_2021, ferreira_galaxy_2024-1}.
We also note that though minor mergers can be difficult to reliably identify, the majority of minor mergers were correctly identified at four or more angles (out of six possible), proving that CNNs can be a path forward in fully understanding the role of all mergers in galaxy evolution.

The lack of a trend in the stellar mass plots for both mergers and nonmergers (Figure~\ref{fig:figure} top center and right) is promising: with the right training data, CNNs enable the identification of less obvious, minor mergers among galaxies with stellar mass $M_\star < 10^{9}M_\odot$.
Important to note for our analysis is that the train, validation, and test sets all include more low-mass than high-mass galaxies and more minor mergers than major mergers. 
This reflects the hierarchical structure that is expected in any 50 Mpc$^{3}$ box of either simulated or real data: far more low-mass galaxies than high-mass galaxies. 
We speculate that when the network incorrectly classifies a high-mass, major merger, it is because it does not see as many examples of these mergers during training.

\section{Conclusions and Outlook}
\label{sec:conclusion}

We used a CNN trained on mock \textit{HST} CANDELS images from the IllustrisTNG simulation at $z \sim 1$ to identify a wide range of galaxy mergers, including masses down to $M_\star = 10^8 M_\odot$ and merger mass ratios down to $\mu = 1/10$. 
The network has a final accuracy of $\sim73\%$.
A few of our highest mass merging galaxies were incorrectly classified, and in the future, we could potentially correct this by combining low-mass galaxies from TNG50 with a sample of high-mass galaxies from the larger simulation box size, TNG100, to create a more balanced and larger training set. 
UMAPs show us that the network is sensitive to the stellar mass and specific star formation rates of the galaxies.
Our nonmerger sample is currently only mass-matched, and with a bigger box size, we could also find SFR-matched nonmergers to break the reliance on SFR. 
By building a training set with similar numbers of major and high-mass mergers as minor and low-mass mergers, we could potentially improve the distinction between mergers and nonmergers for all subcategories of galaxies.

\begin{ack}
A.L.S. and J.M.C. acknowledge support from NASA’s Astrophysics Data Analysis program, grant number 80NSSC21K0646, and NSF AST-1847938.
XS acknowledges the support from the NASA theory grant JWST-AR-04814. 
The work of A.S. was supported by the National Science Foundation MPS-Ascend Postdoctoral Research Fellowship under grant No. 2213288.
L.B. acknowledges support from the NASA Astrophysics Theory program, grant 80NSSC22K0808, and NSF AAG 2307171.
A.\'C: This work was produced by FermiForward Discovery Group, LLC under Contract No. 89243024CSC000002 with the U.S. Department of Energy, Office of Science, Office of High Energy Physics. Publisher acknowledges the U.S. Government license to provide public access under the (\href{http://energy.gov/downloads/doe-public-access-plan}{DOE Public Access Plan}).

We acknowledge the Deep Skies Lab as a community of multi-domain experts and collaborators who’ve facilitated an environment of open discussion, idea generation, and collaboration. This community was important for the development of this project.

\textbf{Author Contributions:} The following authors contributed in different ways to the manuscript.
\noindent Schechter: Writing manuscript, all ML code and analysis.
\noindent \'{C}iprijanovi\'{c}: editing manuscript, mentoring, and overseeing ML code and analysis.
\noindent Shen: Radiative transfer, writing section 2.
\noindent Nevin: Editing manuscript, mentoring, and overseeing ML code and analysis.
\noindent Comerford: Editing manuscript, mentoring, and overall direction of paper.
\noindent Stemo: Assistance with mock image creation.
\noindent Blecha: Initial ideas and direction of paper.

\textbf{Glossary of Symbols:} $z$ redshift; $M_\odot$ solar mass = $2\times10^{30}$kg; $M_\star$ stellar mass; $\mu$ stellar mass ratio $M_{\star, 1}/M_{\star, 2}$; $pc$ parsec =  $3\times10^{16}$m
\end{ack}

%\section*{References}
\bibliography{references}
\bibliographystyle{plain}

%%%%%%%%%%%%%%%%%%%%%%%%%%%%%%%%%%%%%%%%%%%%%%%%%%%%%%%%%%%%

%%%%%%%%%%%%%%%%%%%%%%%%%%%%%%%%%%%%%%%%%%%%%%%%%%%%%%%%%%%%

\newpage
\section*{NeurIPS Paper Checklist}

\begin{enumerate}

\item {\bf Claims}
    \item[] Question: Do the main claims made in the abstract and introduction accurately reflect the paper's contributions and scope?
    \item[] Answer: \answerYes{} % Replace by \answerYes{}, \answerNo{}, or \answerNA{}.
    \item[] Justification: Our abstract includes a few sentences of motivation in the astrophysical field we apply ML to. It additionally includes a statement on the accuracy, purity, and completeness of the model and briefly discusses our interpretive techniques.
    \item[] Guidelines:
    \begin{itemize}
        \item The answer NA means that the abstract and introduction do not include the claims made in the paper.
        \item The abstract and/or introduction should clearly state the claims made, including the contributions made in the paper and important assumptions and limitations. A No or NA answer to this question will not be perceived well by the reviewers. 
        \item The claims made should match theoretical and experimental results, and reflect how much the results can be expected to generalize to other settings. 
        \item It is fine to include aspirational goals as motivation as long as it is clear that these goals are not attained by the paper. 
    \end{itemize}

\item {\bf Limitations}
    \item[] Question: Does the paper discuss the limitations of the work performed by the authors?
    \item[] Answer: \answerYes{} % Replace by \answerYes{}, \answerNo{}, or \answerNA{}.
    \item[] Justification: We discuss limitations of galaxy mass, redshift, and orientation angle in our results section. We additionally discuss potential improvements in our conclusions.
    \item[] Guidelines:
    \begin{itemize}
        \item The answer NA means that the paper has no limitation while the answer No means that the paper has limitations, but those are not discussed in the paper. 
        \item The authors are encouraged to create a separate "Limitations" section in their paper.
        \item The paper should point out any strong assumptions and how robust the results are to violations of these assumptions (e.g., independence assumptions, noiseless settings, model well-specification, asymptotic approximations only holding locally). The authors should reflect on how these assumptions might be violated in practice and what the implications would be.
        \item The authors should reflect on the scope of the claims made, e.g., if the approach was only tested on a few datasets or with a few runs. In general, empirical results often depend on implicit assumptions, which should be articulated.
        \item The authors should reflect on the factors that influence the performance of the approach. For example, a facial recognition algorithm may perform poorly when image resolution is low or images are taken in low lighting. Or a speech-to-text system might not be used reliably to provide closed captions for online lectures because it fails to handle technical jargon.
        \item The authors should discuss the computational efficiency of the proposed algorithms and how they scale with dataset size.
        \item If applicable, the authors should discuss possible limitations of their approach to address problems of privacy and fairness.
        \item While the authors might fear that complete honesty about limitations might be used by reviewers as grounds for rejection, a worse outcome might be that reviewers discover limitations that aren't acknowledged in the paper. The authors should use their best judgment and recognize that individual actions in favor of transparency play an important role in developing norms that preserve the integrity of the community. Reviewers will be specifically instructed to not penalize honesty concerning limitations.
    \end{itemize}

\item {\bf Theory assumptions and proofs}
    \item[] Question: For each theoretical result, does the paper provide the full set of assumptions and a complete (and correct) proof?
    \item[] Answer: \answerNA{} % Replace by \answerYes{}, \answerNo{}, or \answerNA{}.
    \item[] Justification: No theoretical results presented or discussed.
    \item[] Guidelines:
    \begin{itemize}
        \item The answer NA means that the paper does not include theoretical results. 
        \item All the theorems, formulas, and proofs in the paper should be numbered and cross-referenced.
        \item All assumptions should be clearly stated or referenced in the statement of any theorems.
        \item The proofs can either appear in the main paper or the supplemental material, but if they appear in the supplemental material, the authors are encouraged to provide a short proof sketch to provide intuition. 
        \item Inversely, any informal proof provided in the core of the paper should be complemented by formal proofs provided in appendix or supplemental material.
        \item Theorems and Lemmas that the proof relies upon should be properly referenced. 
    \end{itemize}

    \item {\bf Experimental result reproducibility}
    \item[] Question: Does the paper fully disclose all the information needed to reproduce the main experimental results of the paper to the extent that it affects the main claims and/or conclusions of the paper (regardless of whether the code and data are provided or not)?
    \item[] Answer: \answerYes{} % Replace by \answerYes{}, \answerNo{}, or \answerNA{}.
    \item[] Justification: We provide information and citations for all steps taken. All code will be available on GitHub.
    \item[] Guidelines:
    \begin{itemize}
        \item The answer NA means that the paper does not include experiments.
        \item If the paper includes experiments, a No answer to this question will not be perceived well by the reviewers: Making the paper reproducible is important, regardless of whether the code and data are provided or not.
        \item If the contribution is a dataset and/or model, the authors should describe the steps taken to make their results reproducible or verifiable. 
        \item Depending on the contribution, reproducibility can be accomplished in various ways. For example, if the contribution is a novel architecture, describing the architecture fully might suffice, or if the contribution is a specific model and empirical evaluation, it may be necessary to either make it possible for others to replicate the model with the same dataset, or provide access to the model. In general. Releasing code and data is often one good way to accomplish this, but reproducibility can also be provided via detailed instructions for how to replicate the results, access to a hosted model (e.g., in the case of a large language model), releasing of a model checkpoint, or other means that are appropriate to the research performed.
        \item While NeurIPS does not require releasing code, the conference does require all submissions to provide some reasonable avenue for reproducibility, which may depend on the nature of the contribution. For example
        \begin{enumerate}
            \item If the contribution is primarily a new algorithm, the paper should make it clear how to reproduce that algorithm.
            \item If the contribution is primarily a new model architecture, the paper should describe the architecture clearly and fully.
            \item If the contribution is a new model (e.g., a large language model), then there should either be a way to access this model for reproducing the results or a way to reproduce the model (e.g., with an open-source dataset or instructions for how to construct the dataset).
            \item We recognize that reproducibility may be tricky in some cases, in which case authors are welcome to describe the particular way they provide for reproducibility. In the case of closed-source models, it may be that access to the model is limited in some way (e.g., to registered users), but it should be possible for other researchers to have some path to reproducing or verifying the results.
        \end{enumerate}
    \end{itemize}

\item {\bf Open access to data and code}
    \item[] Question: Does the paper provide open access to the data and code, with sufficient instructions to faithfully reproduce the main experimental results, as described in supplemental material?
    \item[] Answer: \answerYes{} % Replace by \answerYes{}, \answerNo{}, or \answerNA{}.
    \item[] Justification: Code will be available on GitHub, and we have a short summary of the compute resources.
    \item[] Guidelines:
    \begin{itemize}
        \item The answer NA means that paper does not include experiments requiring code.
        \item Please see the NeurIPS code and data submission guidelines (\url{https://nips.cc/public/guides/CodeSubmissionPolicy}) for more details.
        \item While we encourage the release of code and data, we understand that this might not be possible, so “No” is an acceptable answer. Papers cannot be rejected simply for not including code, unless this is central to the contribution (e.g., for a new open-source benchmark).
        \item The instructions should contain the exact command and environment needed to run to reproduce the results. See the NeurIPS code and data submission guidelines (\url{https://nips.cc/public/guides/CodeSubmissionPolicy}) for more details.
        \item The authors should provide instructions on data access and preparation, including how to access the raw data, preprocessed data, intermediate data, and generated data, etc.
        \item The authors should provide scripts to reproduce all experimental results for the new proposed method and baselines. If only a subset of experiments are reproducible, they should state which ones are omitted from the script and why.
        \item At submission time, to preserve anonymity, the authors should release anonymized versions (if applicable).
        \item Providing as much information as possible in supplemental material (appended to the paper) is recommended, but including URLs to data and code is permitted.
    \end{itemize}

\item {\bf Experimental setting/details}
    \item[] Question: Does the paper specify all the training and test details (e.g., data splits, hyperparameters, how they were chosen, type of optimizer, etc.) necessary to understand the results?
    \item[] Answer: \answerYes{} % Replace by \answerYes{}, \answerNo{}, or \answerNA{}.
    \item[] Justification: Details in Section 3.
    \item[] Guidelines:
    \begin{itemize}
        \item The answer NA means that the paper does not include experiments.
        \item The experimental setting should be presented in the core of the paper to a level of detail that is necessary to appreciate the results and make sense of them.
        \item The full details can be provided either with the code, in appendix, or as supplemental material.
    \end{itemize}

\item {\bf Experiment statistical significance}
    \item[] Question: Does the paper report error bars suitably and correctly defined or other appropriate information about the statistical significance of the experiments?
    \item[] Answer: \answerYes{} % Replace by \answerYes{}, \answerNo{}, or \answerNA{}.
    \item[] Justification: Error bars are included in the tables and figure, and all details are given in the captions.
    \item[] Guidelines:
    \begin{itemize}
        \item The answer NA means that the paper does not include experiments.
        \item The authors should answer "Yes" if the results are accompanied by error bars, confidence intervals, or statistical significance tests, at least for the experiments that support the main claims of the paper.
        \item The factors of variability that the error bars are capturing should be clearly stated (for example, train/test split, initialization, random drawing of some parameter, or overall run with given experimental conditions).
        \item The method for calculating the error bars should be explained (closed form formula, call to a library function, bootstrap, etc.)
        \item The assumptions made should be given (e.g., Normally distributed errors).
        \item It should be clear whether the error bar is the standard deviation or the standard error of the mean.
        \item It is OK to report 1-sigma error bars, but one should state it. The authors should preferably report a 2-sigma error bar than state that they have a 96\% CI, if the hypothesis of Normality of errors is not verified.
        \item For asymmetric distributions, the authors should be careful not to show in tables or figures symmetric error bars that would yield results that are out of range (e.g. negative error rates).
        \item If error bars are reported in tables or plots, The authors should explain in the text how they were calculated and reference the corresponding figures or tables in the text.
    \end{itemize}

\item {\bf Experiments compute resources}
    \item[] Question: For each experiment, does the paper provide sufficient information on the computer resources (type of compute workers, memory, time of execution) needed to reproduce the experiments?
    \item[] Answer: \answerYes{} % Replace by \answerYes{}, \answerNo{}, or \answerNA{}.
    \item[] Justification: Explanation in Section 3.
    \item[] Guidelines:
    \begin{itemize}
        \item The answer NA means that the paper does not include experiments.
        \item The paper should indicate the type of compute workers CPU or GPU, internal cluster, or cloud provider, including relevant memory and storage.
        \item The paper should provide the amount of compute required for each of the individual experimental runs as well as estimate the total compute. 
        \item The paper should disclose whether the full research project required more compute than the experiments reported in the paper (e.g., preliminary or failed experiments that didn't make it into the paper). 
    \end{itemize}
    
\item {\bf Code of ethics}
    \item[] Question: Does the research conducted in the paper conform, in every respect, with the NeurIPS Code of Ethics \url{https://neurips.cc/public/EthicsGuidelines}?
    \item[] Answer: \answerYes{} % Replace by \answerYes{}, \answerNo{}, or \answerNA{}.
    \item[] Justification: We include appropriate citations for all work, code, and data used here. All authors believe in working in a fair and open work environment.
    \item[] Guidelines:
    \begin{itemize}
        \item The answer NA means that the authors have not reviewed the NeurIPS Code of Ethics.
        \item If the authors answer No, they should explain the special circumstances that require a deviation from the Code of Ethics.
        \item The authors should make sure to preserve anonymity (e.g., if there is a special consideration due to laws or regulations in their jurisdiction).
    \end{itemize}

\item {\bf Broader impacts}
    \item[] Question: Does the paper discuss both potential positive societal impacts and negative societal impacts of the work performed?
    \item[] Answer: \answerNA{} % Replace by \answerYes{}, \answerNo{}, or \answerNA{}.
    \item[] Justification: This work addressed using ML for astronomy, so while ML has larger societal implications, we do not belive the work presented here has social impact.
    \item[] Guidelines:
    \begin{itemize}
        \item The answer NA means that there is no societal impact of the work performed.
        \item If the authors answer NA or No, they should explain why their work has no societal impact or why the paper does not address societal impact.
        \item Examples of negative societal impacts include potential malicious or unintended uses (e.g., disinformation, generating fake profiles, surveillance), fairness considerations (e.g., deployment of technologies that could make decisions that unfairly impact specific groups), privacy considerations, and security considerations.
        \item The conference expects that many papers will be foundational research and not tied to particular applications, let alone deployments. However, if there is a direct path to any negative applications, the authors should point it out. For example, it is legitimate to point out that an improvement in the quality of generative models could be used to generate deepfakes for disinformation. On the other hand, it is not needed to point out that a generic algorithm for optimizing neural networks could enable people to train models that generate Deepfakes faster.
        \item The authors should consider possible harms that could arise when the technology is being used as intended and functioning correctly, harms that could arise when the technology is being used as intended but gives incorrect results, and harms following from (intentional or unintentional) misuse of the technology.
        \item If there are negative societal impacts, the authors could also discuss possible mitigation strategies (e.g., gated release of models, providing defenses in addition to attacks, mechanisms for monitoring misuse, mechanisms to monitor how a system learns from feedback over time, improving the efficiency and accessibility of ML).
    \end{itemize}
    
\item {\bf Safeguards}
    \item[] Question: Does the paper describe safeguards that have been put in place for responsible release of data or models that have a high risk for misuse (e.g., pretrained language models, image generators, or scraped datasets)?
    \item[] Answer: \answerNA{} % Replace by \answerYes{}, \answerNo{}, or \answerNA{}.
    \item[] Justification: Astronomical datasets do not have high risk for misuse.
    \item[] Guidelines:
    \begin{itemize}
        \item The answer NA means that the paper poses no such risks.
        \item Released models that have a high risk for misuse or dual-use should be released with necessary safeguards to allow for controlled use of the model, for example by requiring that users adhere to usage guidelines or restrictions to access the model or implementing safety filters. 
        \item Datasets that have been scraped from the Internet could pose safety risks. The authors should describe how they avoided releasing unsafe images.
        \item We recognize that providing effective safeguards is challenging, and many papers do not require this, but we encourage authors to take this into account and make a best faith effort.
    \end{itemize}

\item {\bf Licenses for existing assets}
    \item[] Question: Are the creators or original owners of assets (e.g., code, data, models), used in the paper, properly credited and are the license and terms of use explicitly mentioned and properly respected?
    \item[] Answer: \answerYes{} % Replace by \answerYes{}, \answerNo{}, or \answerNA{}.
    \item[] Justification: We use data from IllustrisTNG, CANDELS, and Zoobot which is all publicly available and cited, and Zoobot is trained on publicly available SDSS data. 
    \item[] Guidelines:
    \begin{itemize}
        \item The answer NA means that the paper does not use existing assets.
        \item The authors should cite the original paper that produced the code package or dataset.
        \item The authors should state which version of the asset is used and, if possible, include a URL.
        \item The name of the license (e.g., CC-BY 4.0) should be included for each asset.
        \item For scraped data from a particular source (e.g., website), the copyright and terms of service of that source should be provided.
        \item If assets are released, the license, copyright information, and terms of use in the package should be provided. For popular datasets, \url{paperswithcode.com/datasets} has curated licenses for some datasets. Their licensing guide can help determine the license of a dataset.
        \item For existing datasets that are re-packaged, both the original license and the license of the derived asset (if it has changed) should be provided.
        \item If this information is not available online, the authors are encouraged to reach out to the asset's creators.
    \end{itemize}

\item {\bf New assets}
    \item[] Question: Are new assets introduced in the paper well documented and is the documentation provided alongside the assets?
    \item[] Answer: \answerYes{} % Replace by \answerYes{}, \answerNo{}, or \answerNA{}.
    \item[] Justification: Our code to make mock images from publicly available data will be available on a github once approved and in a zip file.
    \item[] Guidelines:
    \begin{itemize}
        \item The answer NA means that the paper does not release new assets.
        \item Researchers should communicate the details of the dataset/code/model as part of their submissions via structured templates. This includes details about training, license, limitations, etc. 
        \item The paper should discuss whether and how consent was obtained from people whose asset is used.
        \item At submission time, remember to anonymize your assets (if applicable). You can either create an anonymized URL or include an anonymized zip file.
    \end{itemize}

\item {\bf Crowdsourcing and research with human subjects}
    \item[] Question: For crowdsourcing experiments and research with human subjects, does the paper include the full text of instructions given to participants and screenshots, if applicable, as well as details about compensation (if any)? 
    \item[] Answer: \answerNA{} % Replace by \answerYes{}, \answerNo{}, or \answerNA{}.
    \item[] Justification: No crowdsourcing or human subjects.
    \item[] Guidelines:
    \begin{itemize}
        \item The answer NA means that the paper does not involve crowdsourcing nor research with human subjects.
        \item Including this information in the supplemental material is fine, but if the main contribution of the paper involves human subjects, then as much detail as possible should be included in the main paper. 
        \item According to the NeurIPS Code of Ethics, workers involved in data collection, curation, or other labor should be paid at least the minimum wage in the country of the data collector. 
    \end{itemize}

\item {\bf Institutional review board (IRB) approvals or equivalent for research with human subjects}
    \item[] Question: Does the paper describe potential risks incurred by study participants, whether such risks were disclosed to the subjects, and whether Institutional Review Board (IRB) approvals (or an equivalent approval/review based on the requirements of your country or institution) were obtained?
    \item[] Answer: \answerNA{} % Replace by \answerYes{}, \answerNo{}, or \answerNA{}.
    \item[] Justification: No human subjects.
    \item[] Guidelines:
    \begin{itemize}
        \item The answer NA means that the paper does not involve crowdsourcing nor research with human subjects.
        \item Depending on the country in which research is conducted, IRB approval (or equivalent) may be required for any human subjects research. If you obtained IRB approval, you should clearly state this in the paper. 
        \item We recognize that the procedures for this may vary significantly between institutions and locations, and we expect authors to adhere to the NeurIPS Code of Ethics and the guidelines for their institution. 
        \item For initial submissions, do not include any information that would break anonymity (if applicable), such as the institution conducting the review.
    \end{itemize}

\item {\bf Declaration of LLM usage}
    \item[] Question: Does the paper describe the usage of LLMs if it is an important, original, or non-standard component of the core methods in this research? Note that if the LLM is used only for writing, editing, or formatting purposes and does not impact the core methodology, scientific rigorousness, or originality of the research, declaration is not required.
    %this research? 
    \item[] Answer: \answerNA{} % Replace by \answerYes{}, \answerNo{}, or \answerNA{}.
    \item[] Justification: LLMs were only used to troubleshoot coding errors and assistance with phrasing in the text.
    \item[] Guidelines:
    \begin{itemize}
        \item The answer NA means that the core method development in this research does not involve LLMs as any important, original, or non-standard components.
        \item Please refer to our LLM policy (\url{https://neurips.cc/Conferences/2025/LLM}) for what should or should not be described.
    \end{itemize}

\end{enumerate}

\end{document}